# Adaptive optics imaging with a pyramid wavefront sensor for visual science

Elisabeth Brunner,[1,*] Julia Shatokhina,[2] Muhammad Faizan Shirazi,[1] Wolfgang Drexler,[1] Rainer Leitgeb,[1] Andreas Pollreisz,[3] Christoph K. Hitzenberger,[1] Ronny Ramlau,[2,4] and Michael Pircher[1] [1]*Center for Medical Physics and Biomedical Engineering, Medical University of Vienna, Waehringer Guertel 18-20, A-1090 Vienna, Austria*
[2]*Johann Radon Institute for Computational and Applied Mathematics, Altenbergerstrasse 69, A-4040 Linz, Austria*
[3]*Department of Ophthalmology and Optometry, Medical University of Vienna, Waehringer Guertel 18-20, A-1090 Vienna, Austria*
[4]*Johannes Kepler University Linz, Industrial Mathematics Institute, Altenbergerstrasse 69, A-4040 Linz, Austria*
*\*Corresponding author: a.elisabeth.brunner@meduniwien.ac.at*

**Abstract**

The pyramid wavefront sensor (P-WFS) has replaced the Shack-Hartmann (SH-) WFS as sensor of choice for high performance adaptive optics (AO) systems in astronomy because of its flexibility in pupil sampling, its dynamic range, and its improved sensitivity in closed-loop application. Usually, a P-WFS requires modulation and high precision optics that lead to high complexity and costs of the sensor. These factors limit the competitiveness of the P-WFS with respect to other WFS devices for AO correction in visual science. Here, we present a cost effective realization of AO correction with a non-modulated P-WFS and apply this technique to human retinal in vivo imaging using optical coherence tomography (OCT). P-WFS based high quality AO imaging was, to the best of our knowledge for the first time, successfully performed in 5 healthy subjects and benchmarked against the performance of conventional SH-WFS based AO. Smallest retinal cells such as central foveal cone photoreceptors are visualized and we observed a better quality of the images recorded with the P-WFS. The robustness and versatility of the sensor is demonstrated in the model eye under various conditions and in vivo by high-resolution imaging of other structures in the retina using standard and extended fields of view.

## 1. INTRODUCTION

With the rise of adaptive optics (AO), the potential of high resolution optical imaging has been unlocked for several applications including astronomy [1], microscopy [2] and ophthalmology [3]. AO enables imaging in or close to the diffraction limit by measuring and correcting wavefront aberrations introduced by propagation of the light through inhomogeneous media or rough interfaces. In ophthalmology imperfections of the eye degrade the retinal imaging quality and AO is used in combination with different imaging modalities such as fundus photography [3], scanning laser ophthalmoscopy (SLO) [4], or optical coherence tomography (OCT) [5, 6] for an in vivo visualization of cellular structures. With the resulting imaging capabilities of these instruments, various cell types such as cone photoreceptors [4, 7], rod photoreceptors [8, 9], retinal pigment epithelium (RPE) cells [10, 11], ganglion cells [12] and more have been identified in the retina from data captured in vivo. These works have revolutionized the imaging options of the living retina, and first commercialisations of AO supported instruments for improved diagnostics and treatment control have been realized.

In classical AO, wavefront aberrations are measured with a wavefront sensor (WFS) and then compensated by a deformable mirror (DM) by adapting its shape. AO systems for retinal imaging commonly use the well-known Shack-

Hartmann wavefront sensor (SH-WFS) [13] which samples the wavefront by the means of an array of lenslets located in a plane conjugated to the pupil of the eye. Non-zero wavefront aberrations cause displacements of the created focal spots at the area detector which provide a measure of the mean wavefront gradients over the corresponding lenslets. While the SH-WFS benefits from the large range in aberration strength over which it has a linear response, the sensor is constrained to a fixed sampling of the wavefront defined by the number of lenslets. Further, the center of gravity computation required for localization of the focal spots is a non-obvious task that is sensitive to intensity variations across the lenslets which for example occur at the borders of the pupil [14, 15].

The pyramid wavefront sensor (P-WFS) [16] has been introduced in astronomical AO where it is steadily supplanting the SH-WFS [17-20]. The main component of a P-WFS is a multi-facet glass pyramid with a large vertex angle. The tip of the pyramid is placed in the system's focal plane and the light is split into as many parts as there are facets. For each part, an image is formed on the detector which is optically conjugated to the pupil plane of the system. In the presence of wavefront aberrations, the images corresponding to each facet show differing intensity distributions. The resulting pupil images are processed as such or combined into two slope-like data vectors [21]. The sensor data is then used to either reconstruct a wavefront estimate [22] which is subsequently mapped on a deformable mirror (DM) or to directly compute the DM actuator commands. Contrary to the SH-WFS, the aberration range for which the sensor has a linear response is small. Therefore, in P-WFS based AO, a modulation is in general applied by either oscillating the beam over the pyramid [23] or by oscillating the pyramid itself [16]. Increasing the radius of the modulation leads to a larger dynamic range of the P-WFS with a trade off in the sensitivity of the sensor [21, 24]. In closed-loop AO correction, the modulation is therefore typically dynamic with a larger radius applied at the start, and a smaller radius once the largest bulk of the aberrations has been corrected. Based on theoretical analysis [25, 26], simulations [27], bench studies [23, 28] and first on-sky results [17, 18] from the astronomical field, a main selling point of the P-WFS is that it shows better sensitivity than the SH-WFS in closed-loop application especially for low order aberration modes. This advantage is enhanced in low light scenarios and for partly corrected wavefronts. These aspects are directly transferable to ophthalmic AO. Next to the above mentioned flexibility in dynamic range and sensitivity, a further advantage of the P-WFS is that the pupil sampling can be easily adjusted [16], to match for example the degrees of freedom of the DM which could be advantageous in woofer- tweeter AO systems [29]. While most AO systems currently in operation at the major ground-based telescopes are based on the SH-WFS, the P-WFS is the sensor of choice for a number of high performance AO systems that will be installed at future extremely large telescopes [19, 20].

In the field of visual science, application of the P-WFS has been proposed in the context of ocular aberrometry [30, 31] and AO aberration correction [32, 33]. With respect to the former, clinical evaluations of the technology have been published and several ocular pyramidal aberrometers are available on the market. In the first works regarding ophthalmic AO, closed-loop correction was achieved with a P-WFS, but no retinal imaging that would allow for an assessment of AO correction quality was performed [32, 33].

In this work, we present the design and implementation of a low-cost 4-sided P-WFS for ophthalmic AO imaging. No modulation is necessary for AO loop convergence which largely simplifies the design of the sensor and increases its applicability for the field of visual science. The performance of the P-WFS was tested in parallel to a SH-WFS based AO-OCT system, where the AO loop is driven by either sensor for comparison. We show that AO-OCT using the P-WFS achieves equivalent or even better performance than using the SH-WFS by visualizing in vivo smallest cone photoreceptor cells that are located in the central fovea of healthy volunteers. Additionally, the P-WFS is successfully tested for various retinal imaging scenarios to demonstrate the flexibility and applicability of the sensor.

## 2. METHODS

We have chosen OCT as imaging modality for the in vivo demonstration of adaptive optics wavefront correction with a non-modulated P-WFS in visual science, but the concept can be directly translated to SLO imaging. The used setup is a lens based spectral domain AO-OCT system as described earlier [11]. In this study, wavefront aberrations introduced by the eye and the system are measured either with the P-WFS or with a customized SH-WFS using a 22 x 22 lenslet array. Wavefront aberration compensation is applied with a DM driven by 69 actuators. The system supports retinal scanning angles of up to 4° x 4° and records volume data of the retina with a 250 kHz A-scan (depth profile) rate. The imaging light has a center wavelength of 840 nm with 50 nm bandwidth which yields a theoretical axial resolution of 4.5 μm in tissue.

**A. Assembly of the P-WFS**

The additional components that are required to implement AO with a P-WFS are depicted in Fig. 1 which shows a schematic of the system and the location of both wavefront sensors. A detailed description of the complete system is given in Supplement 1. In the wavefront sensing path of the system, a 50/50 cube beam splitter is introduced directly in front of the SH-WFS in order to direct half of the light to the non-modulated P-WFS which consists of 2 lenses, a 4-sided glass pyramid and a CMOS detector. With this configuration, potential non-common-path aberrations introduced by the different light pathways to the SH- WFS and the P-WFS, respectively, are reduced to a minimum.

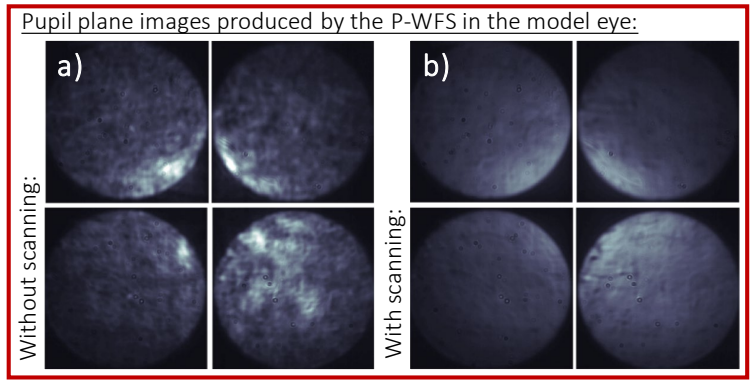

Fig. 2. Cut outs of the four pupil plane images produced by the P-WFS without scanning in a) and with scanning in b). The images were obtained in a model eye under the presence of system aberrations and a slight defocus (overall wavefront error RMS 270 nm, measured with the SH-WFS).

One key feature of the system has proven crucial for the successful implementation of the non-modulated P-WFS for ophthalmic AO: Instead of using a second light source for wavefront sensing, part of the imaging light returning from the retina is used to illuminate both WFSs. This configuration has already been highly beneficial for our SH-WFS based AO [11, 34]. Inhomogeneities in the WFS data that are introduced by the sample structure are averaged out through the fast scanning of the beam over the retina, and reflexes from the lens interfaces and the cornea can be greatly attenuated. Residual reflections from the lens interfaces are removed by polarization optics as only light in a crossed polarization state (in respect to the linear state incident to the eye) is directed to the wavefront sensors [11]. Another aspect of using the imaging light for wavefront sensing is that the size of the focal spot on the retina reduces with the AO correction which improves the wavefront sensing performance. Nevertheless, an implementation of the non-modulated P-WFS with a second light source for wavefront sensing is likely to be successful as well, as long as the WFS light traverses the imaging optics including the DM and the scanners.

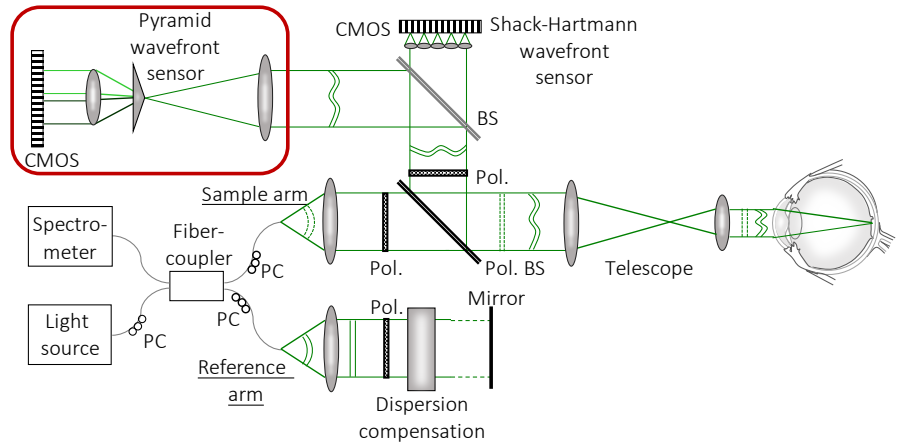

Fig. 1. Partial scheme of the AO-OCT system with a non-modulated P-WFS and a SH-WFS. For simplification only the components essential for wavefront sensing are drawn; deformable mirror and scanners are therefore omitted. The wavefronts of the illumination and of detection light are represented by dashed and solid lines, respectively. PC: polarization controller, Pol.: polarizer, (Pol.) BS: (polarizing) beam splitter.

The assets of this technique do all translate to the P-WFS and most importantly allow to bypass the need for modulation, i.e., applying oscillation of the sensor or the beam, which greatly reduces complexity and costs of the system. In Fig. 2), (pupil) images are shown that were obtained from the P-WFS under the presence of residual system aberrations and slight defocus while using a model eye (which consists of a lens and a scattering surface) as imaging object. The images in Fig. 2.a) were acquired without scanning of the imaging beam across the scattering surface. High frequency intensity variations introduced by the object structure can be clearly observed that degrade the performance of the P-WFS (and are usually removed by sensor modulation). In the presented setup, these are eliminated as soon as the galvanometer scanners are turned on (cf. Fig. 2b), because light returning from different locations of the object reaches the P-WFS during the exposure time of the sensor and as such the influence of the object structure is averaged out. This is an inherent feature of all scanning ophthalmic imaging modalities and there is no need for an additional modulation implementation. Since the wavefront sensing light is part of the imaging beam, the placement of the focal point on the pyramid tip is hereby guaranteed (in scattering objects) even for large scanning angles due to the double pass configuration of the illumination and detection light (see Fig. 1) where the tip of the pyramid lies in a conjugated plane to the tip of the fiber emitting the imaging light. The quality of the P-WFS data greatly improves once the scanners are moving and solely aberration induced intensity variations remain in the pupil images. The P-WFS can be used for closed-loop AO correction with standard calibration techniques using slope like P-WFS data definition (see Supplement 2).

**B. Imaging protocols and subject selection**

The system supports an imaging beam diameter of 7 mm and the AO control algorithm automatically adapts to the observed pupil size of the subject. The presented AO-OCT volume data were recorded with scanning angles of 1° x 1° and 4° x 4° for the small and large field of view (FoV) images, respectively. The sampling densities in x and y direction were chosen with the aim of good visualization of the targeted cellular structures while minimizing motion artifacts. A configuration of 300 x 300 pixels (300 A-scans per B-scan, 300 B-scans) was applied for the small FoV imaging, and a sampling of 750 x 750 pixels was used for the images recorded with a large FoV. In z- (depth) direction the number of pixels is constant with 400 pixels according to the used spectrometer configuration (light spectrum is dispersed over 800 pixels).

The processing applied to the recorded spectral data is similar to standard OCT processing and images are shown on a linear intensity grey scale. For all data volumes, axial displacement between B-scans was corrected in post processing. In case of the large FoV, additionally, the curvature of the retina within the B-scans was compensated for and lateral motion between the B-scans was corrected as outlined in detail elsewhere [11]. All but one (the small FoV images of the anterior retinal layers) of the presented sets of images were extracted from single-shot AO-OCT volumes. This specific image set was retrieved from a data volume that was created by registering and averaging of several image volumes in order to reduce speckle noise [12].

Five healthy subjects ranging in age between 22 and 30 years (mean age of 27.2 years) participated in this study (See Supplement 3). The volunteers were recruited with the selection criteria of negligible media opacities and stable fixation and had a spherical equivalent refraction between <-0.25 and -2.00 Diopters. All measurements adhered to the tenets of the Declaration of Helsinki and were performed after approval of the study by the local ethics committee. Written consent was obtained from each subject prior to the measurement after explaining the nature and form of the procedures. A headrest and a manually adjustable XYZ translation stage were used to stabilize and align the eyes and heads of the subjects. No drugs were administered for dilating the pupil since the pupil diameter under the low light conditions in the lab was larger than 5 mm in all subjects. The volunteers were asked to fixate on an internal fixation target which was optically set to infinity for imaging in the fovea and on an external fixation target for imaging in the periphery. Since accommodation of subjects was not prevented, some data sets showed varying focus settings (associated with varying visibility of the targeted retinal layer) and were discarded.

## 3. EXPERIMENTAL RESULTS

### A. Imaging of the outer retina in the fovea

As a test scenario to demonstrate the repeatability of high performance ophthalmic AO with the P-WFS and to compare the AO imaging quality of the P-WFS and the SH-WFS, we have chosen AO-OCT imaging of the photoreceptor layer at a small FoV of 1° x 1° in the central fovea of 5 healthy subjects with large pupil diameters > 6 mm (see Supplement 3 and 4, respectively). The cell density packing of the cone photoreceptors in the fovea centralis [35] is at the limit of the system's theoretical transverse resolution of ~2 µm. Since the cone photoreceptor density increases exponentially towards the fovea centralis, en-face visualizations of the cone mosaic can serve as an excellent resolution target to judge the performance of ophthalmic AO at a constant pupil size. For the comparison study, several data sets were recorded in a single imaging session alternating between the P-WFS and the SH-WFS, where a pupil sampling comparable to the one provided by the SH-WFS was chosen for the P-WFS. For the small FoV imaging, we opted for the finer OCT sampling of 300 x 300 pixels resulting in a lateral spacing of 1.2 µm. The representative image data displayed in Fig. 3 show en-face images of the photoreceptor mosaic of subject V1_R obtained through depth integration over the outer retinal layers containing the cones and cross-sectional B-scan images extracted at the fovea centralis. The integration range was chosen sufficiently large to include all reflective spots that can be associated with the interface between the inner and outer segments (IS/OS) and with the cone outer segment tips (COST) of single cone photoreceptors. To compare the en-face images obtained with each sensor, power spectra were computed via 2D Fast Fourier transform (FFT) in common regions of interest (RoI) defined as close as possible to the fovea centralis. The radii of the clearly visible Yellott's rings indicate the spatial frequency that scales with the density of the resolved photoreceptor cells. With the P-WFS, single cone photoreceptors can be visualized in the fovea centralis in both the en-face and B-scan image (Fig. 3a + b), clearly matching or even exceeding the image quality achieved with the SH-WFS (Fig. 3c + d). The clearer appearance of the photoreceptor cells manifests itself in the higher clarity of the Yellott's ring obtained for images recorded with AO using the P-WFS. The Yellot's ring is further broader, since a larger amount of cones that are densely packed are visible. A quantitative analysis of the AO imaging quality of the sensors is provided in Supplement 4, followed by a comparison of the WFSs in terms of sensitivity and an assessment of the P-WFS dynamic range, both performed in a model eye.

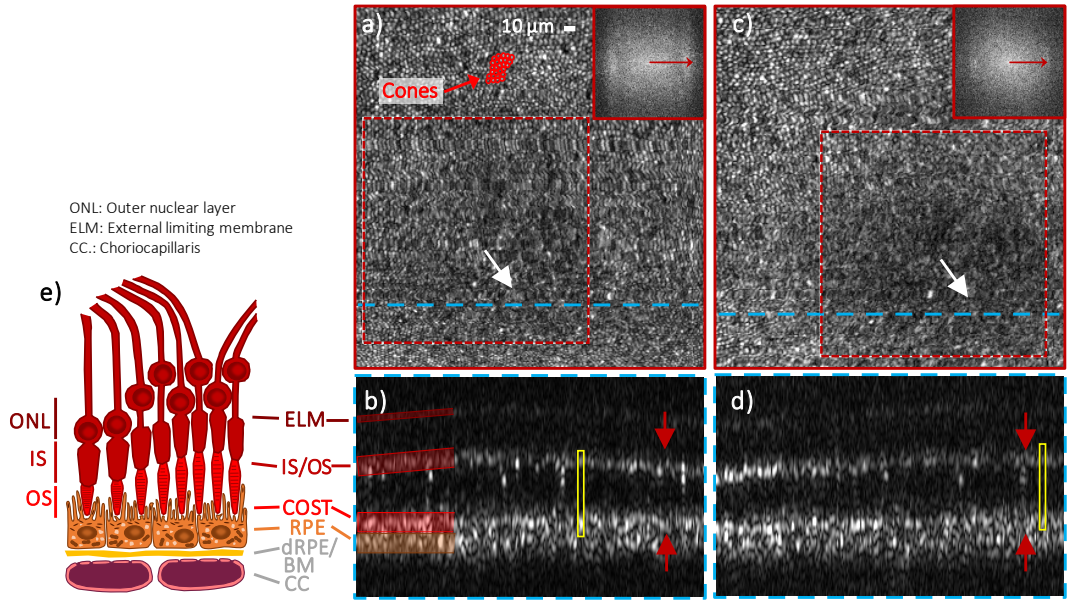

Fig. 3. Comparison of in vivo AO-OCT images of cone photoreceptors obtained with AO correction based on the P-WFS, (a) and b)), and the SH-WFS, (c) and d)). The representative images were retrieved from single data volumes recorded at the fovea of a healthy volunteer (28 years, female, left eye) in one imaging session. The field of view is approximately 0.94° x 0.99°. The en-face projections in a) and c) were created by depth integration over the cone photoreceptor layers and are accompanied by the 2D Fourier transform (FFT) of the marked region of interest which show Yellott's rings. The radii of Yellott's rings indicate the spatial frequency of the cone mosaic that corresponds to the row to row spacing of the cones in the imaged areas. The white arrows point to the same retinal location and indicate the approximate location of the fovea centralis (estimated by the highest density of cones). The blue dashed line highlights the location of the single B-scans shown in b) and d), where the yellow boxes highlight the same cone photoreceptor and the red arrows mark the limits of the en-face integration range. e) is a sketch of the outer retinal bands as known from histology.

In a next step we tested the performance of the P-WFS for larger scanning angles to form a large FoV of 4° x 4°. The A-scan density was adjusted to 750 x 750 pixels in order to obtain good visualization of the cone mosaic pattern while minimizing motion artifacts. Figure 4 shows representative images of the outer retinal layers extracted from a single-shot AO-OCT volume which was recorded in the central fovea of a second subject (V4_L) with a pupil diameter ≈ 7 mm. In the en-face visualization (Fig. 4a), individual photoreceptors can be resolved as close as to a distance of ~0.2° from the fovea centralis. Unlike for the small FoV (see Fig. 3), the photoreceptor cells could not be visualized directly in the fovea centralis when imaging with the large FoV. This can be explained by two factors that are independent of the WFS type: Firstly, the lateral spacing between A-scans is relatively large (~1.9 μm) that leads to a slight under-sampling of the small cone photoreceptors in the fovea centralis. Secondly, despite the optimized system configuration, a compromise is made in AO correction quality when imaging an extended FoV because the wavefront measurements are now averaged over a larger area. The en-face image in Fig. 4a) confirms the large field capabilities of the P-WFS. In the enlarged RoI and the corresponding 2D FFTs, it can be observed how the packing density of the cell mosaic increases with decreasing distance from the fovea centralis which reflects in the larger radii of the Yellott's rings.

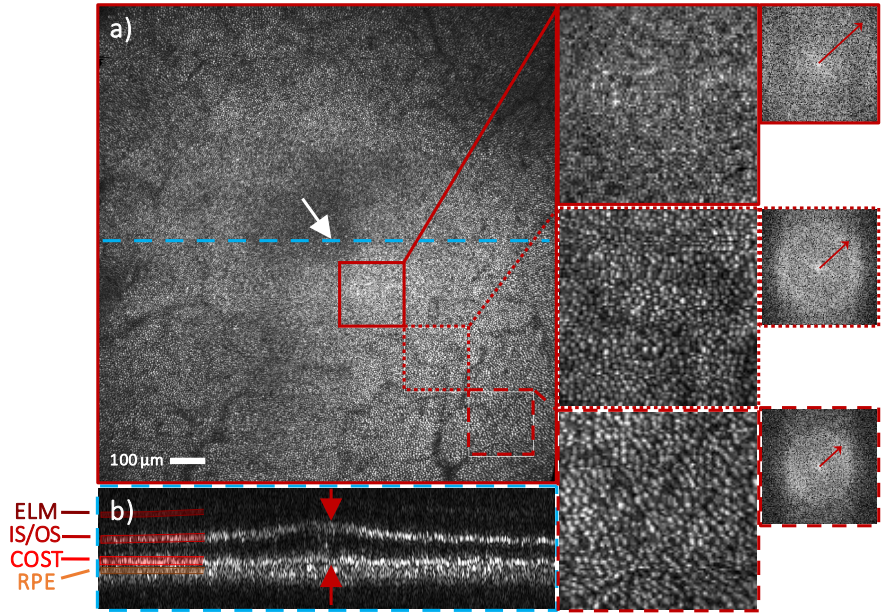

Fig. 4. Representative AO-OCT images recorded in the fovea of a healthy volunteer (29 years, male, left eye) with the P-WFS at an extended field of view of 3.79° x 3.92° (corresponding to 1.36 mm x 1.4 mm on the retina). The en-face image in a) was retrieved from a single data volume by depth integration over the cone photoreceptor layer with the integration range indicated by red arrows in the B-scan image in b). The B-scan location is marked by the blue dashed line in a) where the white arrow points to the approximate location of the fovea centralis. The three en-face images on the right are magnified views of the boxed areas in a). The radii of Yellot's rings in the adjoining 2D FFTs indicate the respective spatial frequencies of the cones at different eccentricities from the fovea centralis.

## B. Imaging of the outer retina in the periphery

In a next step, we recorded AO-OCT data using the P-WFS at an eccentricity of 14° temporal / 6° superior from the fovea of subject V4_R with scanning angles of 1° x 1° and 4° x 4°. Figure 5 displays the representative AO-OCT imaging data. From a single-shot volume, en-face visualizations at several outer retinal layers, a single B-scan showing only the outer retina and an averaged B-scan including the signal from the inner retina were extracted. The en-face visualizations obtained by integration over the IS/OS (Fig. 5a) and COST (Fig. 5b) layers show the coarser packing of the cone photoreceptors at this peripheral imaging location. In the averaged B-scan of Fig. 5g), three bright bands can be clearly distinguished below the COST layer with the top most corresponding to the rod outer segment tips (ROST). In the en-face visualization of the ROST layer (Fig. 5c), the rod photoreceptor cells are observed as hyper-reflective spots ordered around hypo-reflective spots marking the locations of the cone photoreceptors. The false color image (Fig. 5f) is a composite of the en-face images of COST and ROST and visualizes the arrangement of rod photoreceptors that surround cones. The fact that individual rod photoreceptor cells can be resolved, albeit only at a small FoV, in a single-shot AO-OCT volume (see Figs. 5c, f and h) highlights the excellent performance achieved with the P-WFS. The second highly reflective band below the COST layer stems from retinal pigment epithelium cells (RPE), whose mosaic can be clearly visualized in the en-face image of Fig. 5d. Note the different spatial frequencies of RPE cells and cone cells as indicated by the varying radii of Yellott's rings. Posterior to the RPE layer, a third reflective band can be observed which manifests as a granular structure in the en-face visualization and presumptively can be associated with the distal parts [36] of the RPE layer or Bruch's membrane (Fig. 5e).

The AO-OCT image data of the outer retinal bands at the periphery recorded with the large FoV is presented in Supplement 5. With the extended FoV, the different cell mosaics can be visualized over a sufficiently large imaging area to make more conclusive observations about the arrangements of the respective cell types. On the other hand, while the large FoV imaging enables us to see large scale structures like vascularity that are very useful for orientation on the retina, small details might be missed because of the inferior resolution. For example, it can be clearly seen in the small FoV images of Figs. 5d) and h) that the boundaries of the RPE cells which appear as continuous in the large FoV data are in fact formed by discrete points.

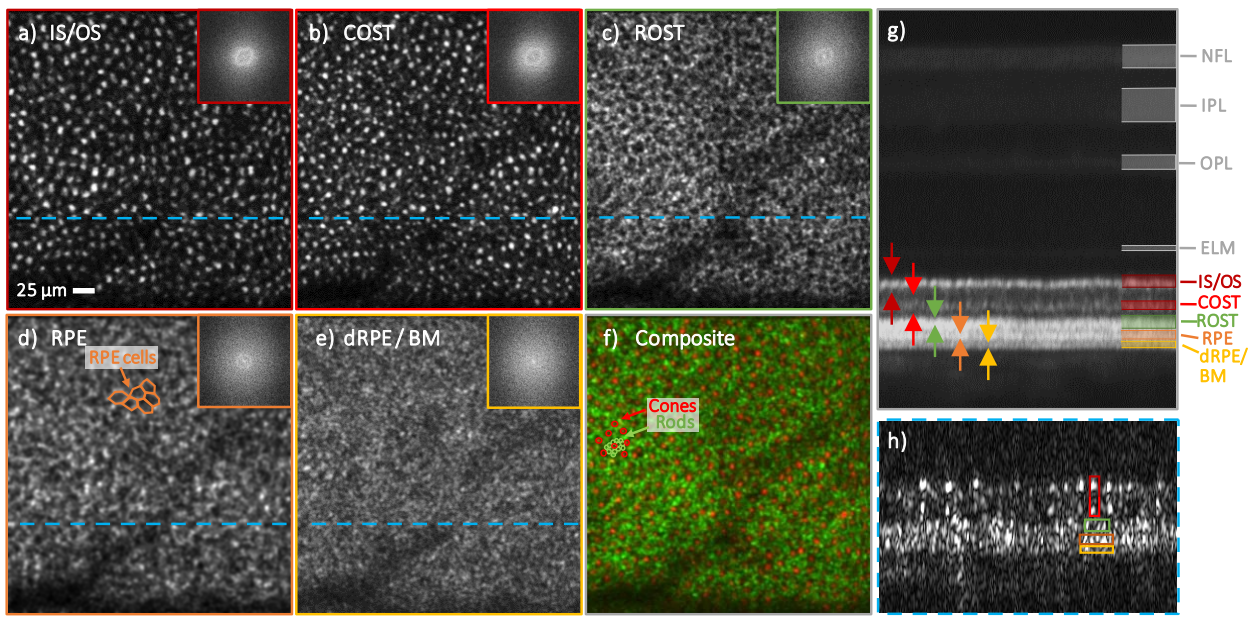

Fig. 5. Different cell types in the outer retinal layer recorded at an eccentricity of 14° temporal / 6° superior in a healthy volunteer (29 years, male, right eye) with a FoV of 0.95° x 1° using AO-OCT with the P-WFS. The en-face images, a) to e), were obtained from a single data volume by integration over different depth ranges in the outer retinal band and have the respective 2D FFTs as an inset. The averaged B-scan in g) shows a projection of the entire volume along the slow scanning direction with the arrows indicating the integration ranges for a) to e). For the following en-face projections, specific retinal cell types can be identified as main contributor in terms of reflected signal: a) junction between inner and outer segments of cone photoreceptors (IS/OS), b) cone outer segment tips (COST), c) rod outer segment tips (ROST), d) retinal pigment epithelium cells (RPE). The structure in e) presumptively corresponds to the distal part of the RPE or the Bruch's membrane (dRPE / BM). The composite in f) is a false color image of COST (red) and ROST (green). In the single B-scan in h), only the outer retinal bands are shown and a selection of the hyper-reflective spots, which form the cell mosaics in a)-d) and the structure visible in e), is marked. The location from which h) was extracted is highlighted in the en-face images with a blue dashed line.

## C. Imaging of the inner retina in the periphery

In a final step we demonstrate the capability of the P-WFS for ophthalmic AO-OCT by imaging inner retinal layers. This requires the application of a specific defocus value to the target wavefront slopes of the P-WFS during AO-correction. The focus of the P-WFS controlled AO was hereby set to the nerve fiber layer (NFL) and AO-OCT data were acquired with small and large scanning angles. For the small FoV imaging of the inner retina, within ~40 sec, a total of 15 data volumes were recorded in three runs each consisting of resetting the deformable mirror, AO loop convergence and consecutive recording of 5 OCT volumes at a scanning angle of 1° x 1° with a 300 x 300 A-scan sampling. Three volumes were corrupted by microsaccades and were therefore discarded. The remaining volumes were registered using a stripe-wise approach and averaged to improve the signal-to-noise ratio and image contrast in the nearly transparent neuronal layers anterior to the photoreceptor bands. From the resulting averaged data volume, only the part in which all the original volumes overlap was used to create the image data presented in Fig. 6. Next to two cross-sectional views, the average of all B-scans (Fig. 6i) and 5 adjacent B-scans (Fig. 6j), en-face images are presented visualizing details in the inner retinal bands: On the surface of the inner limiting membrane (ILM), irregular highly reflective structures can be observed (Fig. 6a) that were associated with macrophage-like cells [37] or microglial cells [38]. In the nerve fiber layer (NFL), single nerve fiber bundles crossed by a large arteriole/venule are visualized (Fig. 6b). In the ganglion cell layer (GCL), ganglion cell somas of different sizes (8 – 24 µm) can be distinguished in the lateral view in Fig. 6c) as well as in the cross-section in Fig. 6j). The relatively small number of AO-OCT volumes (compared to previous work [12]) that were required to visualize ganglion cells attests the high AO imaging performance and sensitivity achieved with the P-WFS based system. Further, en-face images were extracted from four depth locations within and at the outer edge of the inner plexiform layer (IPL) (Fig. 6d-g), which show dense meshes of high spatial frequency irregularities which have been reported in the literature as dendrites and synapses between GCs, amacrine cells, and bipolar cells [12]. Despite the focus being set to the NFL, the resolution and image contrast in the en-face image in Fig. 6h) is sufficient to visualize details in this transverse slice located in the vicinity of the outer plexiform layer (OPL). Representative imaging data of the inner retina recorded with the P-WFS and an extended FoV of 4° x 4° is shown in Supplement 5.

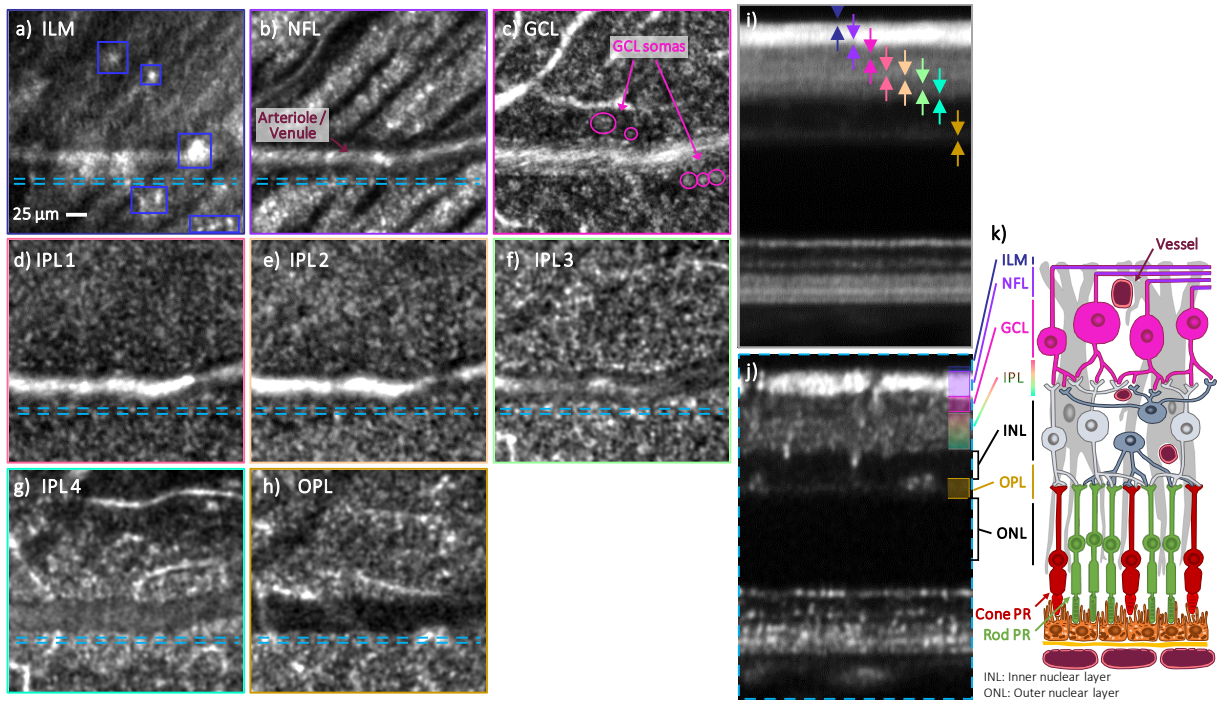

Fig. 6. Different structures of the inner the retina visualized for a healthy volunteer (29 years, male, right eye) at 14° temporal / 6° superior with AO-OCT using the P-WFS. The focus was set to the nerve fiber layer and the FoV is 0.82° x 0.78°. The representative images are extracted from a volume obtained by averaging 12 registered volumes that were recorded within ~40 sec. The en-face images a)-h) were obtained by depth integration over parts of or in vicinity of the following retinal layers: a) inner limiting membrane (ILM), b) nerve fiber layer (NFL), c) ganglion cell layer (GCL), d)-g) inner plexiform layer (IPL), h) outer plexiform layer (OPL). The integration ranges are indicated by color-coded arrows in the averaged B-scan in i) which was obtained by projection of the full data volume in the slow imaging direction. The dashed lines in the en-face images mark the location of the 5 adjacent B-scans averaged in j) in the data volume. k) is a sketch of the retinal layers and cells as known from histology.

## 4. DISCUSSION AND CONCLUSION

In this work we demonstrate high performance of a non-modulated pyramid wavefront sensor (P-WFS) to drive an adaptive optics (AO) system in visual science. The configuration was tested in vivo in 5 healthy volunteers and stable closed-loop AO correction was achieved reliably. The obtained excellent imaging quality using AO-OCT for imaging is due to the high quality of the optical system as well as the better performance of the P-WFS in comparison with the used SH-WFS. It should be mentioned that the subjects' pupils were not artificially dilated and accommodation was not suppressed leading to the necessity of discarding of some volume data sets and a slight degradation in resolution in subjects with smaller pupil size. The performance of the ophthalmic AO imaging with the P-WFS was demonstrated in various retinal imaging scenarios of interest by displaying en-face and cross-sectional views extracted from AO-OCT volumes. Specifically, the visualization of rods and central foveal cones underlines the performance of AO correction provided by the sensor. The applicability of the sensor was tested by imaging in the central fovea and in the periphery (14° temporal / 6° superior) as well as by shifting the focus of the imaging beam to the inner retinal layers. The sensor is usable for both, small (1° x 1°) and large FoV imaging (4° x 4°) indicating a similar variety of applications as is known for the SH-sensor.

AO correction was achieved in the eye with a standard calibration procedure based on a slope-like P-WFS data definition. Oscillation of the pyramid or modulation of the beam, which is generally applied for the P-WFS to increase the linear regime of the sensor response and subsequently the dynamic range of the sensor was not necessary which greatly simplifies the overall design of the sensor. While it is not compulsory to use the P-WFS only in the linear regime if the sensor signals are used to drive closed-loop AO correction, we hypothesize that the excellent and stable AO performance was achieved because the sensor response is to a certain extent linearized by using part of the imaging light for wavefront sensing and the scanning of the imaging beam over a larger area on the retina during P-WFS exposure. As aberrations will be corrected for both, the illumination and detection light paths, the size of the focal intensity distribution (point spread function) at the retina is changed during AO-correction. This intensity distribution, however, serves as source for wavefront sensing. It has been shown that using an extended light source instead of an ideally point-like source for wavefront sensing creates a similar effect as beam oscillation [27, 30]. Larger aberrations, in the eye dominated by defocus and astigmatism, lead to a broadening of the focal intensity distribution at the retina and subsequently at the tip of the pyramid. During convergence of the AO loop, aberrations are corrected for, leading to a sharper focal intensity distribution at the retina and at the sensor, approximating the condition of a point-like source for wavefront sensing. This can be viewed similar to a dynamic beam modulation which is generally adopted to create a large dynamic range with large modulation at the beginning of the correction, and a high sensitivity with small modulation once the loop is close to convergence.

The second aspect is that the shape of the focal intensity distribution at the tip of the pyramid is not only determined by wavefront aberrations but by tissue structure (scattering potential of the illuminated area on the retina) as well. During exposure of the P-WFS the beam moves over a larger area of the retina (on its way back to the sensor the beam is de-scanned), thus averaging out the influence of the underlying tissue structure on the shape of the focal intensity distribution at the pyramid. This shape is therefore solely determined by aberrations that are introduced to the light beam on the return path from the retina.

The presented images of central foveal cones recorded with both sensors show better results achieved with the P-WFS in comparison with the SH-WFS when using comparable pupil sampling. At the current stage this cannot be generalized and requires further experiments for confirmation. A major asset of the P-WFS is that it provides flexibility in terms of pupil sampling which can be changed optically or by digitally binning the pixels. It is therefore possible to provide optimal pupil sampling for different types of wavefront correctors without changing the main hardware component of the WFS. The pupil sampling chosen in this work was unnecessarily high for operating a deformable mirror with 69 actuators, but we could see several advantages in the flexible pupil sampling as compared to the rigid sampling of the SH-WFS. Typically, in a SH-WFS, partly illuminated lenslets at edges of the pupil negatively affect the wavefront sensing [15, 39]. Therefore, it is common practice to discard these in the calibration of the SH-WFS. We observed that the missing data points led to a slight instability in the control of the outer actuators of the DM. For the P-WFS calibration, only a small number of rows of digitally binned pixels with reduced intensity were excluded at the pupil border. Due to the higher sampling of the P-WFS, we did not observe any negative effect on the outer actuators since a proportionally smaller number of sampling points had to be excluded.

During in vivo imaging, the subjects show differing pupil sizes and shapes which are often smaller than the system pupil size. In the current implementation, the sub-optimal but frequently implemented approach [14] of filling missing SH-WFS and P-WFS data points with zero was adopted. The higher pupil sampling of the P-WFS naturally reduces artifacts introduced by this approach [15]. Next to irregular pupils, eye movements pose another challenge to AO imaging, especially in clinical applications. Pupil tracking based on SH wavefront sensing data has been proposed previously [40] and could be performed with higher precision using the densely sampled pupil images provided by the P-WFS before binning and without affecting the AO-correction performance. Finally, by implementing a zoom optical relay it is possible to change the P-WFS pupil sampling in a continuous manner [16]. This would allow for a WFS based study of various spatial frequencies present in ocular wavefront aberrations and might give insight into the optimal wavefront corrector required for the application. While it is possible to model the low order aberrations present in the eye, simulation of higher order aberrations as they are, e.g., introduced by the tear film have been harder to predict.

In terms of temporal dynamics, studies have shown that correction bandwidths of ~30 Hz are required to sufficiently correct lower order ocular wavefront aberrations [41, 42]. Higher bandwidths are essential to follow fast temporal dynamics of higher order aberrations but this requires that more light is dedicated for wavefront sensing which comes at the cost of lower imaging sensitivity. Our initial results obtained in a model eye confirmed studies from astronomical AO [25, 27] which suggest a better performance of the P-WFS in comparison with the SH-WFS in the case of a low light scenario which might be beneficial for improving the AO bandwidth while maintaining high imaging sensitivity. An in-depth comparison of the sensors for the context of ophthalmic imaging needs to be subject of future studies. Depending on the outcome of these studies, the P-WFS might on the long term replace the SH-WFS in visual science similar to what can be observed in astronomy.

For the presented P-WFS, a standard, low-cost glass pyramid was used. Since for the P-WFS the detection of the wavefront sensing light is not at a focused but a diffused plane, the requirements on the dynamic range of the cameras can be eased, further adding to the low-cost character of the suggested WFS. Finally, we want to point out that the P-WFS concept presented in this paper can be directly translated to confocal scanning laser microscopy for sensor based AO correction.

## ACKNOWLEDGEMENTS

This project has received funding from the Austrian Science Fund (SFB Tomography across the scales F6803-N36 and F6805-N36) and the European Commission (MERLIN-H2020 ICT 780989). The authors want to thank S. Esposito, L. Busoni and M. Bonaglia (Astrophysical Observatory of Arcetri, Italy), R. Biasi (Microgate, Italy) and V. Hutterer (Industrial Mathematics Institute Linz, Austria) for fruitful discussion. Special thanks go to the Astrophysical Observatory of Arcetri for lending us the pyramid used in this work. Further the authors thank S. Steiner and A. Sedova (Medical University of Vienna, Austria) for the clinical measurements.

# Supplementary material

## 1. AO-OCT SYSTEM, ASSEMBLY AND ALIGNMENT OF THE P-WFS

A scheme of the used spectral domain AO optical coherence tomography (OCT) is depicted in Fig. S1 and is based on a system that was described earlier [1]. In the wavefront sensing path a non-polarizing 50:50 beam splitter is placed that allows aberration measurements either with the non-modulated pyramid wavefront sensor (P-WFS) or with the customized Shack-Hartmann wavefront sensor (SH-WFS). The compensation of the wavefront aberrations is performed with a 69-actuator deformable mirror (DM) from the company ALPAO (Montbonnot, France). Wavefront sensing is performed using the light back-scattered from the retina that is in an orthogonal polarization state compared to the linear polarization state incident to the eye [1]. OCT imaging is done with a spectrometer that consists of a collimator, a diffraction grating, a focusing lens, and a line scan camera (spL4096-140 km, Basler, Ahrensburg, Germany).

The newly designed non-modulated P-WFS is assembled from a four faceted glass pyramid with a large vertex angle, two achromatic lenses and a detector. The first lens (Lens1, f1 = 100 mm) is placed at a distance corresponding to one focal length from a conjugated pupil plane of the system (cf. blue dashed line in the wavefront sensing arm in Fig. S1). The sensing light is focused at the tip of a glass pyramid (custom made) which has a vertex angle of 160° and the dimensions 25.4 mm × 25.4 mm ×7.2 mm. Unlike in P-WFS based astronomical AO, no high precision component was used, but the pyramid was produced with standard optical manufacturing processes resulting in the following accuracies. The roof has a width of ~120 μm and the edges of ~15 μm. The cost of the pyramid is 500 € per unit. The vertex of the pyramid is positioned in the back focal plane of the first lens (yellow dashed line in Fig. S1). Due to the double-pass configuration of the imaging and detection light, the pyramid vertex is optically conjugated to the fiber tip emitting the imaging light and to the focal spot created on the retina. The theoretical width of the point spread function at the pyramid tip is ~4.4 μm, calculated at the sensing wavelength of 840 nm. The pyramid splits the field into four parts and introduces four different tilts, angularly separating the four beams. A second lens (Lens2, f2 = 30 mm) is used to conjugate the plane of the detector, a CMOS camera (Photonfocus QR1 - D2048x1088-384-G2-8), to the pupil. The diameter of the four pupil images is ~2.2 mm and each contains up to 391 samples across the diameter according to the amount of digital pixel binning that is applied. A custom graphic user interface written under Labview (National Instruments, Austin, USA) and MATLAB (The Mathworks, Inc., Natick, USA) processes the pupil images.

The customized SH-WFS was described in detail earlier [2]. It consists of a lenslet array with 32 x 32 micro-lenses (Adaptive Optics Associates C-0300-16-S) that is located in a pupil plane of the system. The focal spots formed by the lenslets are recorded with a CMOS camera (Pixelink PL-A741). The focal length of the lenslet array is 16 mm and the lenslet pitch is 300 μm. A commercial adaptive optics software, Alpao Core Engine, running in MATLAB is used to compute the centroids of the SH-WFS focal spots which provide a sampling of 22 across the pupil diameter. The specifications of both sensors are summarized in Table S1.

During preliminary tests of the implemented non-modulated P-WSF, we observed that very exact alignment of the P-WFS components was crucial to obtain the AO correction performance required for the visualization of smallest retinal cells such as central foveal cone photoreceptors. Both lenses and the detector were therefore mounted on mechanical stages that allow for lateral and axial adjustments in the micrometer range. The pyramid is attached to the stage of the reimaging lens Lens2 by means of a cage system in which the pyramid can be manually rotated and moved along the optical axis independently from the lens. The alignment was performed with a flat mirror inserted between the first telescope and the deformable mirror to minimize the effect of systems aberrations on the alignment procedure. The mirror was adjusted such that the light was coupled back into the single mode fiber (which ensures that the light beam is back-reflected in itself).

**Table S1. Specifications of the Shack-Hartmann and the pyramid wavefront sensors**

|  | P-WFS | SH-WFS |
|---|---|---|
| Pixel size of the detector | 5.5 μm | 6.7 μm |
| Dimension of the detector | 2040 x 1088 pixels | 1180 x 1024 pixels |
| Bit depth of the detector | 8 | 10 |
| Sampling of the pupil plane (considered in the experiment) | flexible (digitally binned pixels) | 22 x 22 subapertures |

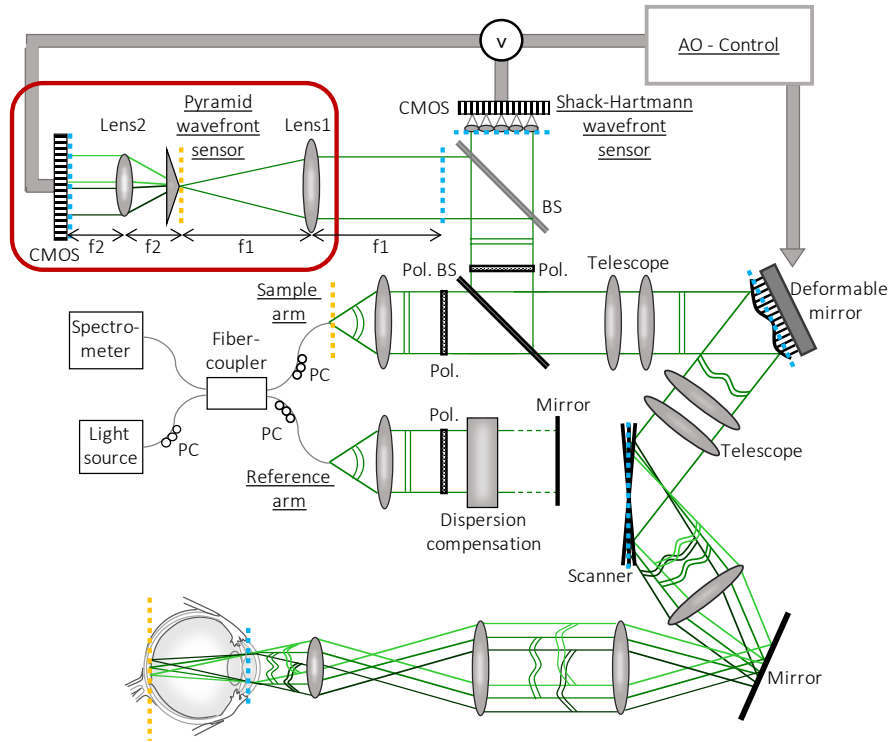

Fig. S1. Scheme of the AO-OCT system with a non-modulated P-WFS and a SH-WFS. For simplification only one galvanometer scanner and only 2 of the 4 beams created by the P-WFS are drawn. PC: polarization controller, Pol.: Polarizer, (Pol.) BS: (polarizing) beam splitter. The dashed blue and yellow lines mark pupil and focal planes of the system, respectively.

In the first step of the P-WFS assembly, the CMOS camera was placed in the WFS arm orthogonally to the collimated beam at a distance behind the pupil plane in the P-WFS branch which corresponds to the 4f configuration depicted in Fig. S1. The center pixel of the detector was laterally aligned to the peak of the Gaussian beam profile. Subsequently, Lens1 was mounted at distance f1 behind the pupil plane and adjusted laterally to place the peak of the diverging beam profile on the center pixel of the detector. Then, Lens2 was added at distance f1+f2 behind the first lens and laterally aligned in the same manner. We verified that the peak of the beam profile did not change position on the detector when Lens2 is axially moved, which confirmed that the lenses were placed orthogonally to the beam. In order to guarantee exact axial alignment of the reimaging lens, the beam was diverted by means of a flat mirror inserted between Lens2 and the detector. The axial position of Lens2 was then fine-tuned until collimation of the diverted beam could be observed over a distance of 5 m. Before the pyramid could be introduced, the axial position of the camera had to be optimized in order to achieve conjugation of the detector to the pupil plane of the system. For this purpose, Lens1 was removed from its mount and the axial position of the camera was adjusted such that the diameter of the focal spot created in the detector by Lens2 was minimized. The pyramid was then mounted into the cage system which is laterally centered with Lens2, while Lens1 remained removed from the assembly. By minimizing the diameter of the four focal spots appearing at the detector, the axial position of the pyramid was determined and rotational alignment was achieved by placing the focal spots in a square with the sides parallel to the axes of the detector.

For the fine tuning of the axial and lateral alignment of the pyramid, Lens1 was reinstalled in the system and a model eye (which consists of a lens and a scattering surface) placed at the eye's pupil location was used as imaging object (the flat mirror between the first telescope and the DM was removed). During detector exposure the galvanometer scanners were turned on and the imaging beam was moving over the surface with a small scanning angle. Since the P-WFS was implemented as an add-on to an existing complete system, we chose to correct for system aberrations during alignment to avoid the possibility of a negative influence of the system aberrations on the final AO performance of the P-WFS. It is however important to mention that the P-WFS could also be aligned in a separate setup without system aberrations using a defined point light source and then be added to the system (as was done for the SH-WFS). This corrective shape of the deformable mirror was computed with the SH-WFS, but it is also possible to correct for systems aberrations with a sensor-less approach [3, 4]. The stage which holds both Lens2 and the pyramid was now moved laterally until the intensity power was equally distributed among the four pupil images produced by the P-WFS and axially until the intensity was distributed as uniformly as possible within the pupil images.

## 2. P-WFS DATA PIPELINE AND CALIBRATION

In Fig. S2, the customized P-WFS data pipeline is illustrated with pupil images obtained in vivo for subject S1. The sensor data is computed from the four reimaged pupils recorded on the CMOS camera (cf. Fig. S2a).

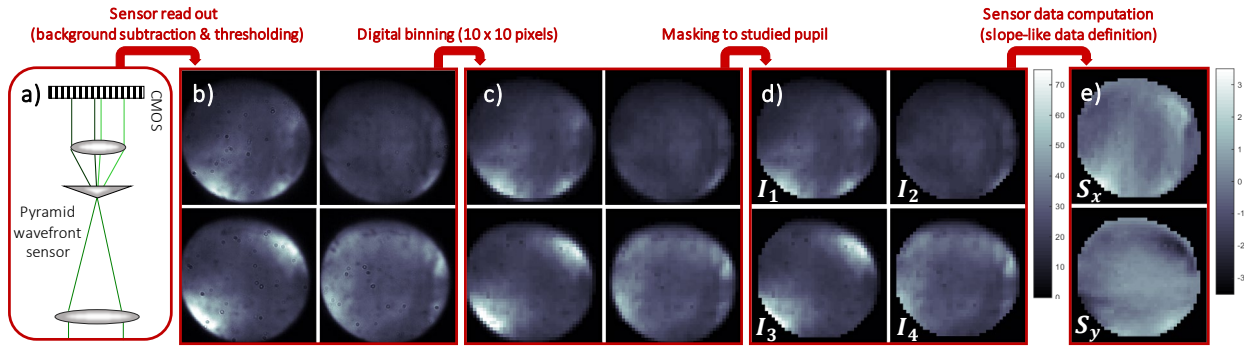

Fig. S2. Data pipeline implemented for the non-modulated P-WFS depicted in a): Cut outs of the four pupil plane images obtained from the P-WFS after background subtraction and intensity thresholding are shown in b), followed by the output of the digital binning routine in c). Only the illuminated pixels within the studied pupil in d) are used according to the pupil image numbering for the computation of the two data maps in e) via the standard approach of slope-like P-WFS data definition. The data was obtained in vivo.

The software reads out a pixel area of 1080 x 1080 pixels with each pupil image containing 410 pixels in diameter. After subtraction of a pre-recorded background illumination frame and application of an intensity threshold (cf. Fig. S2b), digital binning is applied via pixel value averaging (cf. Fig. S2c). For the binning of pixel groups at the borders of the pupil which are partly located inside and partly outside of the threshold mask, only pixels within the threshold mask are averaged in order to avoid introduction of errors at the edges of the pupil. For the presented AO data, several amounts of pixel binning were applied resulting in between 13 and 391 binned pixels across the pupil diameter. During in vivo AO imaging, the eye pupil may change in size or move laterally. The effective pupil in each AO correction step is therefore defined as intersection between the studied pupil (pre-defined in the calibration procedure) and the pupil defined by the intensity thresholding. The resulting four slightly different pupil images (cf. Fig. S2d) can be either used directly (full-frame approach), or are combined into two slope-like data maps. Thorough theoretical analyses of the forward model of the pyramid sensor [5, 6] provide explicit mathematical expressions connecting the sensor data with the corresponding incoming wavefront. In this work the standard slope-like P-WFS data definition is applied and the P-WFS "slopes" $S_x$ and $S_y$ are computed at each point $(x,y)$ of the effective pupil using the following formulas [7]:

$$S_x(x,y) = \frac{1}{I_0}\left[\left(I_1(x,y) + I_4(x,y)\right) - \left(I_2(x,y) + I_3(x,y)\right)\right], \tag{S1}$$

$$S_y(x,y) = \frac{1}{I_0}\left[\left(I_1(x,y) + I_2(x,y)\right) - \left(I_3(x,y) + I_4(x,y)\right)\right], \tag{S2}$$

where $I_i(x,y)$ is the intensity at point $(x,y)$ in pupil image $i$ and the pupil images at the CMOS camera are numbered according to Fig. S2d. A normalization is applied by $I_0$ which is the mean intensity in the four pupil images where only the effective pupil is considered. For small aberrations, the normalized intensity differences in Eqs. (S1) and (S2) are proportional to the vertical and horizontal wavefront gradients. The P-WFS data points within the studied pupil that are not illuminated are set to zero. The resulting P-WFS data maps contain also information about the observed tip and tilt modes which do not have to be corrected due to the double-pass configuration of the imaging and detection light. Tip and tilt can be filtered from the P-WFS data by subtracting, respectively, the mean value of $S_x(x,y)$ and $S_y(x,y)$ computed over the effective pupil from the slope maps, providing the final P-WFS data maps shown in Fig. S2e).

The DM actuator commands required for compensating the ocular aberrations are calculated directly from the P-WFS data through a zonal control approach [8]. The system was calibrated in the model eye for closed-loop operation while the galvanometer scanners were moving the beam over the object with a small scanning angle of 0.5°. P-WFS data maps were acquired for the poking action of each DM actuator. The pseudo-inverse of the resulting interaction matrix was obtained via a truncated singular value decomposition [9] and used as AO control matrix for computing the closed-loop commands of the DM actuators. In the presented calibration procedure, we assume a linear relationship between actuator commands and sensor responses. As mentioned in the introduction of the main manuscript, the P-WFS shows linear behavior for only very small wavefront aberrations and generally dynamic modulation is applied to increase the dynamic range of the sensor [7, 10, 11]. In this work, AO correction with the P-WFS is achieved without beam modulation. It has to be subject of future investigations, to which extent the averaging applied by the scanning of the imaging beam across the retina increases the linear range of the P-WFS similar as it has been shown for beam modulation. In order to account for potential non-linearity, we performed the calibration close to zero aberrations which minimizes the error in linearization of a nonlinear process. For this purpose, an actuator command vector correcting for system aberrations was added to the bias command vector of the deformable mirror. This pre-correction was computed with the SH-WFS of the system, but could alternatively be obtained through a sensor-less approach. The studied pupil is now defined based on the recorded pupil images after intensity thresholding and the desired amount of digital binning. It has proven beneficial for the AO performance achieved with this P-WFS implementation to reduce the resulting mask by a small number of outer circles of binned pixels. Before applying the actuator pokes, a reference frame was recorded with the P-WFS and two reference data maps were computed according to the data pipeline described above (cf. Fig. S3a). Each actuator was set to 20% of the maximum stroke while the other actuators stayed in their bias position and response maps were obtained from the recorded pupil plane images. The reference maps were subtracted from the poking responses of the P-WFS and the resulting relative P-WFS poking responses were stored in the interaction matrix. An example of the absolute and relative response of the P-WFS to poking of DM actuator Nr. 25 are shown in Figs. S3b) and S3c). At a later stage of this work, we performed the P-WFS calibration without pre-correction of the system aberrations and found that the resulting in vivo AO imaging performance was comparable to the quality achieved with a pre-correction in the calibration step. This preliminary result suggests that the P-WFS calibration is robust to small system aberrations.

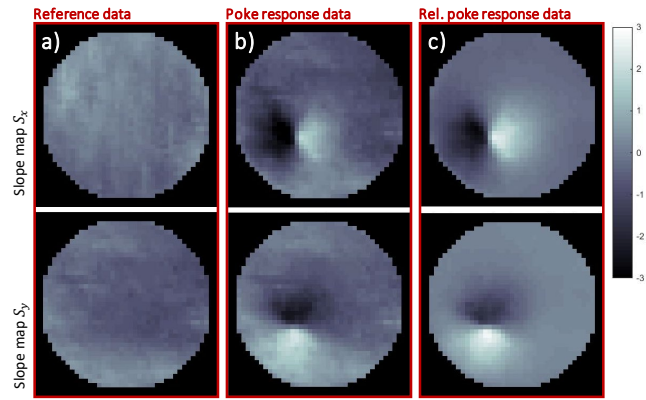

Fig. S3. P-WFS calibration data for computation of the sensor response to 20% stroke at actuator Nr. 25: a) Reference sensor data maps, b) absolute sensor response maps and c) relative sensor response maps.

For calibration of the SH-WFS we adapted the procedure from [2] to match the above mentioned steps for the P-WFS. Instead of using the original calibration of the SH-WFS for closed-loop correction, that had been performed with a flat mirror at the eye pupil location and without scanner motion, the calibration was performed in the model eye, with moving scanners and with the same pre-correction applied (to minimize system aberrations). The same zonal control approach as for the P-WFS using the same number of controllable modes was applied for AO correction. Since after the pre-correction of the system aberrations, the SH-WFS sees a plane wavefront there is no need for the acquisition of a new reference frame. Note here a difference to the P-WFS, which in addition sees the non-common path aberrations between the two sensors (even if those are small), and therefore requires a respective reference frame. For construction of the SH-WFS interaction matrix, a centroid map was acquired for each actuator poke and translated to two wavefront gradient maps using the Alpao Core Engine. The SH-WFS control matrix is also obtained via computation of the pseudo-inverse of the interaction matrix.

The AO software for closed-loop correction is configured to enable switching between the input from the P-WFS and the SH-WFS to drive the AO control and the DM actuator command updates are computed by multiplying the gradient map measured for the residual wavefront with the respective control matrix obtained in the calibration. No pre-correction of the system aberrations is applied and the control iteratively adjusts the actuators such that the measured gradient maps converge towards zero correcting for all aberrations introduced by the eye and the system. The illuminated pupil is defined via intensity threshold in each AO correction step and might not cover the entire studied pupil considered in the calibration because of eye motion or smaller pupil diameters. For both sensors, the slope values at sample points that lie outside the illuminated pupil are set to zero. A constant loop-gain of 0.25 and an exposure time of 40 ms was applied for both sensors for all in vivo measurements. The AO control software has not yet been optimized for speed leading to a correction bandwidth of ~6 Hz and the AO loop currently converges within a few seconds. To ensure that the entire field of view (FoV) is covered within a single exposure of the WFS, the sampling (number of B-scans) in slow direction is reduced during loop convergence. The measured wavefront aberrations are therefore averaged over the entire FoV and only aberrations that do not change over the FoV are corrected. During recording of the AO-OCT volumes, the sampling in the slow direction is reset to full resolution and the AO correction keeps running. Therefore, the wavefront measurements are now averaged over the full range of the fast scanning direction and only over a part of the slow scanning direction.

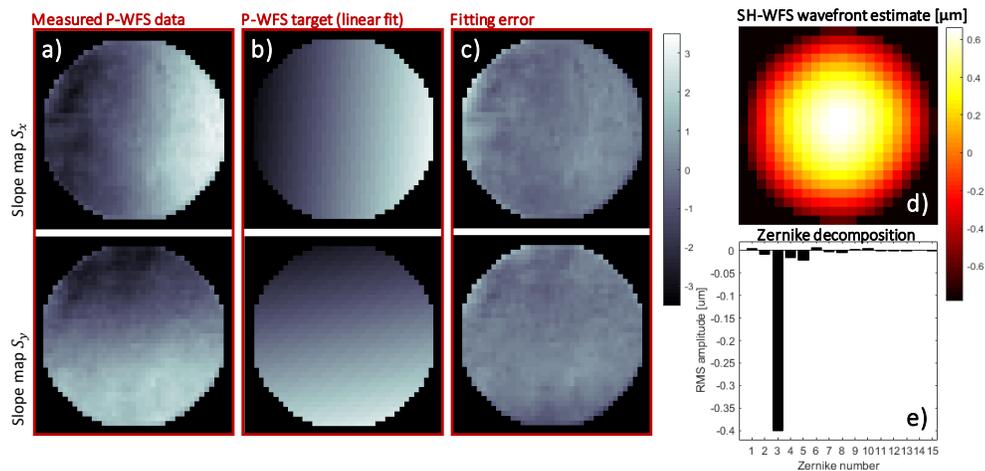

Fig. S4. Target computation for closed-loop focus shifting with the P-WFS: The P-WFS gradient maps measured for a defocus wavefront of 0.4 μm root mean square (RMS) amplitude are shown in a), followed by the target gradient maps computed through a linear polynomial fit of the measured data in b) and the polynomial fitting error in c). The defocus wavefront was measured with the SH-WFS, where d) shows the wavefront estimate and e) the corresponding Zernike decomposition (including the first 15 modes according to the Noll definition [12], except piston).

An important feature of any AO system for retinal imaging is the capability to set the focus of the imaging beam to distinct layers in the retina as the depth of focus is considerably smaller than the depth extension of the retina. In the used system, only light returning from the retina that is in an orthogonal polarization state (with respect to the incident polarization state) is directed to the wavefront sensors. As the back-scattered light from the RPE is in a random polarization state [13], light originating from this layer will mainly contribute to the wavefront measurement. Thus, the inherent focus of the system at hand is on the outer retinal layers [1]. In order to shift the focus to the anterior layers of the retina, the AO control must drive the sensor data maps not towards zero but towards data maps that represent a defocus wavefront. For the P-WFS, such target sensor data maps were computed via a polynomial fit from sensor data recorded in the model eye for different amounts of defocus which were introduced by displacement of the artificial retina after pre-correction of system aberrations. The P-WFS data maps in Fig. S4a) were measured for a defocus wavefront of 0.4 µm root mean square (RMS) amplitude. A wavefront estimate obtained with the SH-WFS and the corresponding Zernike decomposition of the estimate in Figs. S4d) and e) confirm that the dominant component in the present aberration was defocus. The linear surface fits to the measured P-WFS data maps, displayed in Fig. S4b), were used as target for focus shifting with the P-WFS. The polynomial fitting error maps in Fig. S4c) show the components of the measured P-WFS data which were discarded by the linear fit. The low order variations can be attributed to the residual astigmatism modes seen in the Zernike decomposition of the reconstructed wavefront and the variations of higher spatial frequency in the error maps can be explained by the effect of the pupil edges and high order residual aberrations. The presented closed-loop AO routine for inner retina focusing with the P-WFS is a preliminary implementation and in its current form less stable than the routine for outer retina focusing. Therefore, the following procedure is applied: first the wavefront aberrations are corrected using zero maps as target, then the target is switched to the fitted defocus sensor data maps and the loop is stopped once the signal from the anterior layers is sufficiently strong which can be checked via the real time cross sectional view of the retina (OCT-B-scan). Then the data recording is started.

## 3. P-WFS BASED AO FOR IMAGING HEALTHY VOLUNTEERS

Table S2 gives a summary of the volunteers' characteristics. The volunteers were chosen with a natural pupil diameter between 6 mm and 7 mm in dark environment and no drugs for artificial pupil dilation were administered. The pupil diameters along the x and y axis of the pupil plane were estimated from the P-WFS pupil images before digital binning and reveal not only variations in size but also in the shape of the pupils. The RMS amplitudes of the wavefront aberrations were measured with the SH-WFS. The wavefront reconstruction was performed with the Alpao Core Engine using a modal approach [2]: Accommodation was not prevented during the measurements which lead to a slight variation in the measured RMS amplitudes of the wavefront estimates. Therefore, the means of the measured wavefront amplitudes are provided. The standard deviation in the values stayed < 0.7 µm for all considered eyes. The RMS wavefront amplitudes measured for the volunteers extended over a considerable range from ~0.3 µm to almost 1 µm. All subjects showed significant to strong high order aberrations. Volunteer V5 was wearing a contact lens for both the wavefront aberration measurement and the volume recording. Additionally, a routine clinical eye examination including eye length measurements with the ZEISS IOLMaster 500 (Carl Zeiss Meditec AG, Jena, Germany) was performed for each volunteer before the AO-OCT measurement. The mismatch for the Diopter (measured with a Refrakto Keratometer ARK1-S, NIDEK, Gamagōri, Japan) and wavefront aberration RMS values in some subjects is due to the non-linearity of open-loop wavefront measurements with the SH-WFS for larger Diopter values. The imaging data presented in the main manuscript were recorded with volunteers V1 and V4.

**Table S2. Characteristics of healthy volunteers included in the study**

| Volunteer | | V1 | V2 | V3 | V4 | V5* |
|---|---|---|---|---|---|---|
| Age [years] | | 28 | 30 | 27 | 29 | 22 |
| Gender | | Female | Male | Female | Male | Male |
| LE: | Pupil diameter [mm] | 6.69 x 6.74 | 6.14 x 6.16 | 6.05 x 6.44 | 6.91 x 6.83 | / |
| | Diopters (Sphe Asm) | -0.25 + 0.25 | -2.00 + 2.50 | -1.75 + 0.75 | < 0.25 + 0.50 | |
| | Wavefront aberration [µm RMS] | 0.496 | 0.801 | 0.941 | 0.439 | |
| RE: | Pupil diameter [mm] | 6.91 x 6.91 | 6.48 x 6.13 | 6.05 x 6.30 | 6.74 x 6.65 | 6.14 x 6.21 |
| | Diopters (Sphe Asm) | -0.50 + 0.25 | -1.50 + 2.00 | -1.25 + 0.25 | < 0.25 + 0.00 | < 0.25 + 0.00 |
| | Wavefront aberration [µm RMS] | 0.752 | 0.792 | 0.936 | 0.556 | 0.350 |
| Explanations: LE refers to the left eye and RE to right eye of the subject. In the main text, the eye is indicated by _L and _R, respectively. If marked with *, the subject was wearing a contact lens. | | | | | | |

In Fig. S5, we provide representative AO-OCT image data recorded in the fovea of 5 healthy subjects to demonstrate the applicability of the P-WFS for AO imaging in more subjects. For the P-WFS based AO correction, a pupil sampling of 37 x 37 (10 x 10 pixel binning) was applied. The data was acquired at a small field of view FoV

(scanning angles of 1° x 1°) and a OCT sampling of 300 x 300 A-scans. All images were obtained from single-shot volumes, no volume averaging was applied. Note that Yellott's rings (insets in Fig. S5) are more blurred along the slow imaging direction because eye motion mainly distorts structures in this direction.

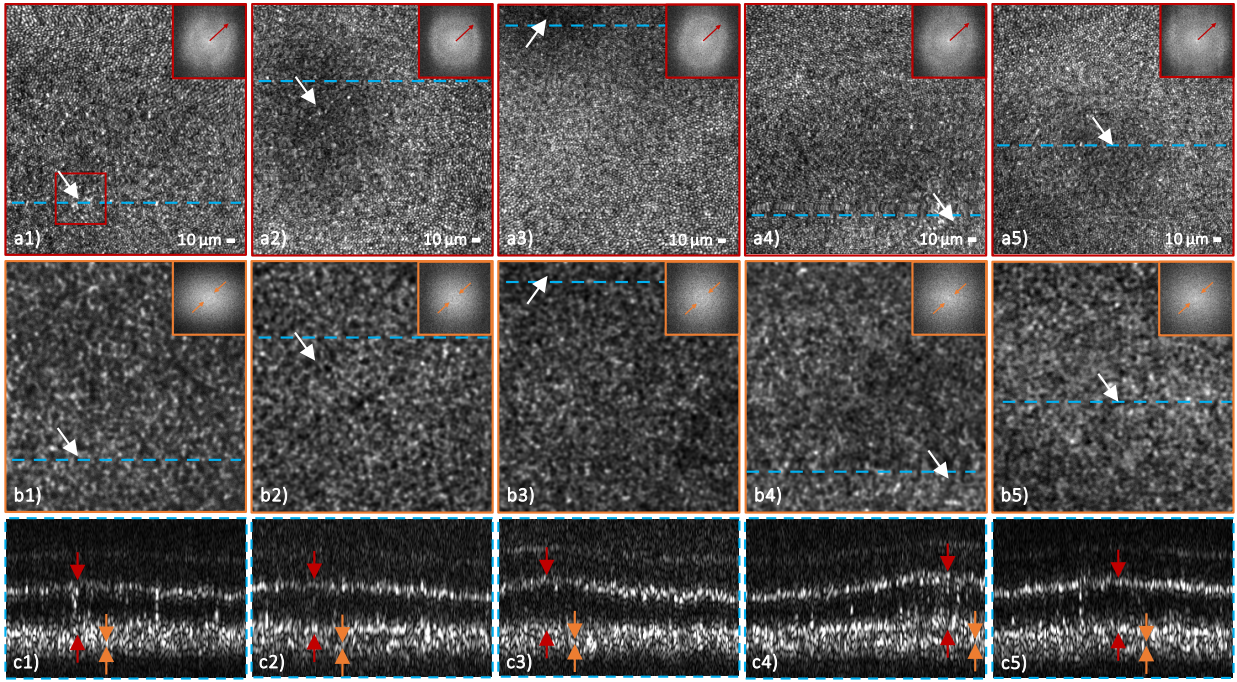

Fig. S5. Representative AO-OCT images recorded in the central fovea of 5 healthy volunteers with the P-WFS and a field of view of 0.94° x 0.99°. The numbers in the image labels indicate the subject numbers that are also shown Table S2 (data sets 1-3 and 5 were recorded in the right eye and data set 4 in the left eye of the respective subject). The en-face images in a1)-a5) and b1)-b5) were retrieved from single data volumes by depth integration over the cone photoreceptor layers and the layer of retinal pigment epithelial cells (RPE), respectively. The radii of Yellot's rings in the corresponding 2D Fourier transforms (FFT) indicate the spatial frequencies of the cone and RPE mosaics that correspond to the row to row spacing of the cones and RPE cells in the imaged areas. The blue dashed lines highlight the locations of the single B-scans shown in c1) -c5), which were extracted in the fovea centralis The white arrows point to the approximate location of the fovea centralis (estimated by the highest density of cones). The red and orange arrows mark the limits of the en-face integration ranges.

For volunteers V1 and V4, the photoreceptors could even be resolved in the most central part of the fovea centralis where the cones are most densely packed (cf. Figs. S5a1 and S5a4). The larger pupil diameter sizes of V1 and V4 yield a better theoretical lateral resolution but also increase the impact of wavefront aberrations on the retinal images. The visualization of the most central foveal cone photoreceptors is therefore a telltale sign of an excellent AO correction quality obtained with the P-WFS. In all presented cross-sectional views, the cone photoreceptor layers do not appear as continuous bands but as rows of discrete points corresponding to the junctions of IS/OS and COST of single cone photoreceptors. In the layer of the cone OS, more sparsely distributed highly reflective spots can be observed. In the en-face images of the RPE (Figs. S5b1 – S5b5), the characteristic mosaic of RPE cells can be clearly identified. The hexagonal shapes of the RPE cells are formed by hyper-reflective discrete points which most likely originate from scattering organelles in the RPE cells [14]. Direct imaging of RPE cells remains a challenge because of their low intrinsic contrast and typically extensive averaging of several volumes is used to enhance cell contrast taking advantage of cell motility [14]. We previously demonstrated single volume AO-OCT imaging of RPE cells at a large FoV of 4° x 4°with the SH-WFS [1]. The extended FoV improves the visibility of the Yellott's ring since more RPE cells are visualized and contribute to the generation of the ring. In this work, faint Yellott's ring could be visualized for single volume AO-OCT images of the foveal RPE layer recorded at a small FoV of 1° x 1° highlighting the exceptional AO performance achieved by the P-WFS. The Yellott's rings in the power spectra of the RPE mosaics have smaller radii than for the cone photoreceptor mosaics due to the larger row to row spacing of the RPE cells. Further, their appearance is sharper since the row to row spacing of the RPE cells is constant throughout the field of view. The spacing of the foveal RPE cells was quantified for all volunteers by computing the radial average of the respective power spectra. The determined row to row spacings (14.1 µm, 13.7 µm, 12.7 µm, 12.4 µm and 12.6 µm for subjects V1, V2, V3, V4 and V5, respectively) are in the expected range [1, 14]. The density of the cone photoreceptor packing increases exponentially towards the fovea centralis and a range of spatial frequencies are observed resulting in a broader Yellott's ring for the photoreceptor layers. For subject V1, the row to row spacing of the cone photoreceptors in the fovea centralis was obtained in the same manner from the power spectrum computed for the area (0.2° x 0.2°) marked in the en-face image of Fig. S5a1. The resulting row to row spacing of 2.65 µm, which (assuming a hexagonal cone packing) corresponds to a next neighbor spacing of 3.06 µm, is in good agreement with literature [15, 16].

## 4. COMPARISON OF THE P-WFS AND THE SH-WFS IN VIVO AND IN A MODEL EYE

For the in vivo comparison of the AO imaging quality provided by our non-modulated P-WFS and the SH-WFS, AO-OCT images of the cone photoreceptor mosaic at the central fovea were recorded at a small FoV of ~ 1° x 1° using both sensors in a single imaging session. The experiment was repeated in 8 eyes (both eyes of volunteers V1 to V4). By changing the digital binning, the number of samples used by the P-WFS was adjusted to a pupil sampling of 21 x 21 which is comparable to the 22 x 22 sampling used by the SH-WFS.

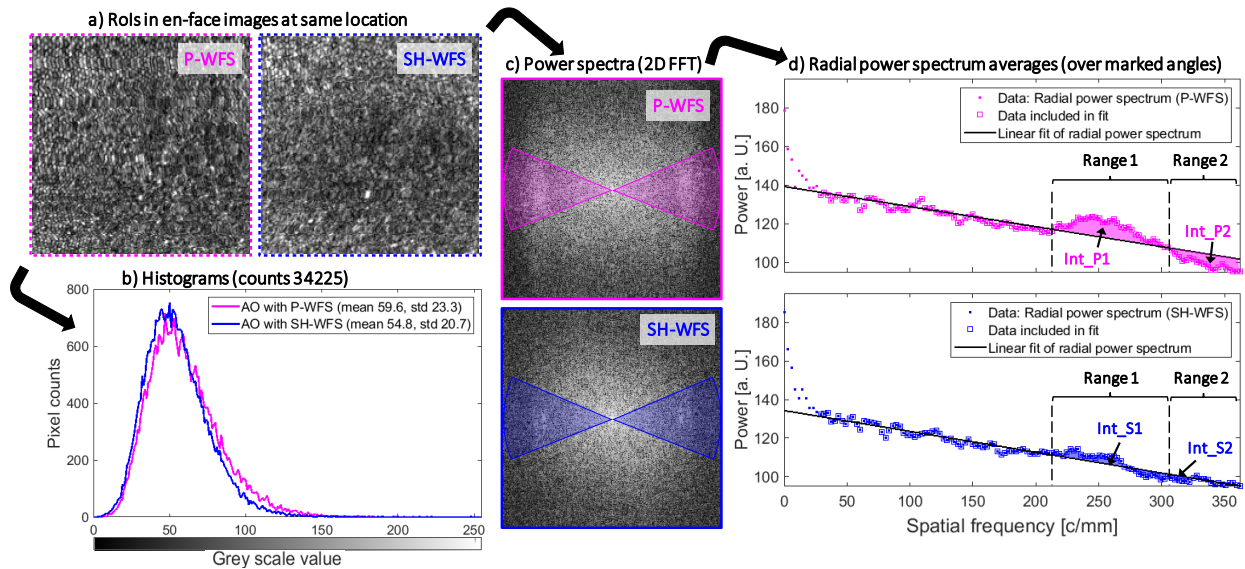

Fig. S6. Quantitative comparison of the P-WFS and the SH-WFS by retrieving AO imaging performance metrics from en-face AO-OCT images of the cone photoreceptor mosaic recorded in the central fovea of healthy volunteers: a) shows the common regions of interest (RoIs) defined as close as possible to the fovea centralis. For the RoIs, histograms in b) and power spectra (2D FFT) in c) are obtained. The power spectra are radially averaged over the angles marked in c) and fitted with a linear polynomial in d). The final performance metrics are the standard deviation of the RoIs provided by the histograms and the values obtained by integrating the difference between the radial power spectrum averages and the respective fits over the spatial frequency ranges Range 1 and Range 2.

In order to avoid any possible bias in favor of the P-WFS regarding tear film degradation or decrease of the pupil size due to an extended stay in a dark environment (i.e. loss in resolution), we first recorded the data with the SH-WFS. To account for motion artifacts or small misalignments of the subject, and thus minimize their influence on the comparison, the following imaging protocol was used: For each sensor, the loop was opened and closed 3 times. After every loop convergence, 5 AO-OCT volumes were recorded in a row, resulting in a total of 15 AO-OCT volumes per sensor. The AO loop kept running during acquisition. One entire imaging session covering both sensors took ~ 2min. Volumes affected by motion, micro-saccades or accommodation were discarded, resulting for all considered eyes in at least 5 remaining volumes for each sensor that were of identical aberration correction quality, respectively. From the remaining valid data sets, en-face images showing the cone mosaic were created and identical regions of interest

(RoIs) of 0.6° x 0.6°, located in or as close as possible to the fovea centralis, were chosen in both en-face images (see marked areas in Fig. 3a and c). For these RoIs the power spectra were computed via 2D-FFT and compared. For each considered eye, the imaging quality in all valid volumes obtained with the P-WFS was better than the imaging quality of all volumes obtained with the SH-WFS (clearer appearance of photoreceptors as well as of the Yellott's rings). In a next step, we performed a quantitative analysis of the AO imaging data obtained with the P-WFS and the SH-WFS in terms of imaging contrast and spatial frequency information via observing the radial average of the Yellott's rings in the corresponding power spectra. In Fig. S6, the procedure to obtain the chosen performance metrics from the RoIs (Fig. S6.a) and the corresponding power spectra (Fig. S6.c) is visualized (same data sets of V1_R as in Section 3.A of the main manuscript). For the depiction in Fig. S6, the contrast was adjusted in the RoI images and the power spectra. The quantitative analysis was performed with the unaltered data. Histograms of the grey scale values (0-256) were computed for the RoIs (Fig. S6.b) which show higher standard deviation (and thus contrast) for the en-face image obtained with the P-WFS than for the en-face image obtained with the SH-WFS. In the power spectra (Fig. S6.c), the Yellott's rings are washed out along the slow scanning axis due to motion. To account for this, the computation of the radial averages of the power spectra was performed only for the angles marked by the cones in Fig. S6.c).

It can be clearly seen in Fig. S6d), that, for the P-WFS data, power is shifted from the high spatial frequency range (> 310 c/mm) to the spatial frequency range which corresponds to the expected range of cone photoreceptor row-to-row spacing. This power shift is significantly less pronounced for the SH-WFS data, suggesting that the smaller cone photoreceptors are not resolved and instead washed out into noise of high spatial frequency. In order to quantify this power shift, linear polynomial fits were performed for the two radially averaged power spectra. For the different subjects, the RoIs were always located in the central fovea but at slightly different eccentricities. The improved imaging resolution obtained with the P-WFS resulted in clear Yellott's rings for all subjects. The spatial frequencies present in the imaged cone mosaic sections could be identified by the cross-over points of the radial power spectrum averages and the corresponding linear fits (see dashed line in top plot of Fig. S6d). These cross-over points create two spatial frequency ranges for each subject: Range 1, corresponding to the spatial frequencies observed in the cone mosaic resolved with the P-WFS, and Range 2, corresponding to higher spatial frequencies. As a metric for the observed spatial frequencies, integration of the differences between the averaged radial power spectra and the corresponding linear fits was performed over Range 1 and Range 2 for the P-WFS and the SH-WFS data, resulting in the values (Int_P1, Int_P2) and (Int_S1, Int_S2), respectively. Values of Int_P1 and Int_S1 far above zero and values of Int_P2 and Int_S2 far below zero indicate a clear Yellott's ring and therefore excellent AO imaging quality for the respective data set.

Table S3. Quantitative comparison of the P-WFS and the SH-WFS (all metrics given in a. U.)

| Volunteer | V1_L | V1_R | V2_L | V2_R | V3_L | V3_R | V4_L | V4_R |
|---|---|---|---|---|---|---|---|---|
| Hist. (P): (mean/std) | (64.2/**27.4**) | (59.6/**23.3**) | (73.8/**31.1**) | (69.6/**32.5**) | (71.4/**26.8**) | (60.7/22.7) | (59.0/**24.2**) | (58.2/24.4) |
| Hist. (SH): (mean/std) | (55.9/22.4) | (45.8/20.7) | (78.4/28.0) | (88.9/31.2) | (53.6/20.8) | (81.7/**30.7**) | (63.8/23.8) | (83.9/**30.8**) |
| P. Sp. (P): (Int_P1, Int_P2) | **(544/-319)** | **(497/-287)** | **(894/-489)** | **(829/-454)** | (46/-28) | **(406/-224)** | **(373/-207)** | **(497/-287)** |
| P.SP. (SH): (Int_S1, Int_S2) | (346/-198) | (108/-37) | (1/-30) | (160/-86) | (**80**/**-39**) | (66/-38) | (175/-122) | (-4/0) |

In Table S3, the quantitative comparison of the P-WFS and SH-WFS AO imaging quality is summarized for all considered subjects with the superior value for each metric printed in bold and the eye in which the data was recorded indicated by _L and _R. The table shows that the P-WFS outperforms the SH-WFS for 6 out of 8 eyes in terms of image contrast (std of the grey values of the image). For the measure of resolved spatial frequency information, the P-WFS showed highly superior values in 7 out of 8 eyes. The low values obtained for subject V3_L indicate an overall poorer visibility of Yellott's ring and the difference between the sensors lies within the error margin. Overall, the quantitative assessment of the AO imaging quality confirms a superiority of the P-WFS to the SH-WFS in our system.

To highlight the sampling flexibility of the P-WFS we performed AO imaging (in V1_R and V2_R) when using different pupil samplings and compared the results with data obtained with the SH-WFS. The different pupil samplings were realized digitally by changing the amount of digital binning. The following 4 pupil samplings were tested in a single imaging session: 391 x 391 (no digital binning); 37 x 37, 21 x 21, 13 x 13. The AO loop converged in a stable manner and comparable P-WFS based AO imaging quality was reliably achieved with all samplings. We can therefore conclude that in the presented configuration, the P-WFS shows high performance independent of the pupil sampling. Considering AO correction bandwidth, we however do not recommend to use the settings where no binning is applied. From application in astronomical AO, it is known that oversampling of the P-WFS–allows to mitigate possible performance loss due to pupil position misalignments [17]. We therefore used the 37 x 37 pupil sampling option for the recording of the main part of the presented AO-OCT data (cf. Figs. 4-6 and Figs. S5, S8 and S9). The 21 x 21 pupil sampling is comparable to the pupil sampling provided by the SH-WFS and was used for the comparison study presented above. The sampling of 13 x 13 is the lowest sampling with which a sufficient number of sampling points is obtained to achieve high resolution images with a 69 actuator DM [18]. While these initial results provide some insight on the flexibility of the P-WFS in terms of pupil sampling, a more thorough study is required to draw final conclusions on the optimal pupil sampling for ophthalmic imaging.

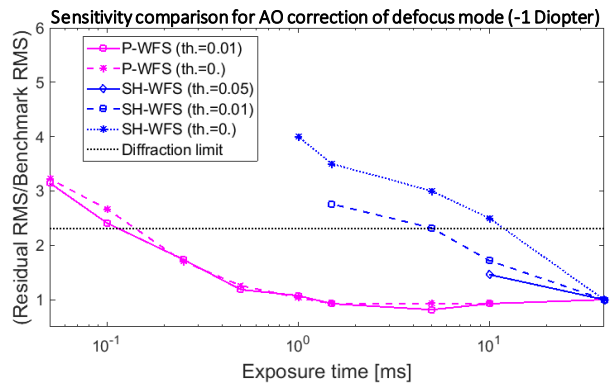

Fig. S7. Comparison of the P-WFS and SH-WFS sensitivity. The residual wavefront RMS error after AO loop convergence (measured with the SH-WFS at 40ms exposure time) is plotted for different exposure times. Several intensity thresholds are considered for the center of gravity computation of the SH-WFS focal spots and the P-WFS pupil image masking. The benchmark RMS value corresponds to the residual wavefront RMS value after AO loop convergence obtained for the respective sensor setting at 40 ms exposure time. The dashed horizontal line marks the performance required for diffraction limited imaging according to the Marechal criterion.

Next to the quantitative assessment of the in vivo AO imaging quality, we further performed a comparison of the P-WFS and SH-WFS in terms of sensitivity and an assessment of the P-WFS dynamic range in a model eye. The model eye consists of a lens with 30 mm focal length and a scattering surface which can be axially moved, serving as artificial retina. It was verified that the light power incident at both sensors is equal and the pupil sampling of the P-WFS matched the pupil sampling of the SH-WFS. Since the artificial retina shows stronger scattering than the human retina, the light power was reduced at the source to obtain similar power in the WFS arm as in an in vivo imaging scenario. Defocus aberrations were introduced by shifting the artificial retina out of the lens' focal plane that are then corrected by the deformable mirror driven either by the P-WFS or the SH-WFS.

The sensitivity provided by each WFS was assessed by correcting for a constant level of defocus (1 Diopter) at different exposure times ranging from 0.025 ms to 40 ms. The residual wavefront RMS value after loop convergence was then measured using the SH-WFS with a 40 ms exposure time. In Fig. 7, the ratio between the obtained residual RMS values and a benchmark, which consists of the residual RMS value obtained for the respective sensor setting at 40 ms exposure time, is plotted for the different exposure times. An AO correction is considered as failed if the residual RMS value is larger than 2 times the diffraction limited RMS value (computed with the Marechal criterion). Different levels of intensity thresholding were considered for the center of gravity (CoG) computation of the SH-WFS focal spots and for the masking of the P-WFS pupil images before applying digital binning. We observed that changes in the CoG threshold highly impacted the SH-WFS behavior for reduced exposure times, while the intensity thresholding on the P-WFS data did only marginally affect the AO correction performance achieved for each exposure time level. For intensity thresholds larger than zero, the SH-WFS provided diffraction limited AO correction down to 5 ms exposure time. The P-WFS was more robust to low flux scenarios and showed no significant loss in AO correction performance down to 0.5 ms exposure for both threshold settings. Loop instabilities due to discarded WFS samples and a slowdown in loop convergence were observed for exposure times <10 ms and < 0.25 ms for the SH-WFS and the P-WFS, respectively. Applying a zero threshold resulted in a highly unstable AO loop for the SH-WFS, while it had no such effect on the P-WFS performance. A zero threshold is not recommendable for an in-vivo application of the P-WFS because changing pupil sizes have to be taken into account. We conclude that in our system the gain in limiting exposure time of the P-WFS is at least of one order of magnitude if compared to the SH-WFS. It should be noted that different CMOS detectors are used, however the detector area dedicated to the computation of each pupil sample is of the same order of magnitude for both sensors. An explanation for the different behaviors of the sensors for small exposure times possibly lies in the underlying principle of the sensors which are of very distinctive nature. A more detailed analysis of the matter is beyond the scope of this work and should take into account other low order aberrations and high order aberration, as well as the impact of the CMOS camera specifications on these results.

We further added an assessment of the P-WFS dynamic range by stepwise increasing the defocus aberrations and observing the residual wavefront after convergence of the AO correction. The exposure time was set to 40 ms and the intensity threshold to 0.01. Myopic and hyperopic eyes were modelled by introducing negative or positive defocus aberrations through shifting the artificial retina. Within the range of -3.5 Diopter to 3.5 Diopter, the P-WFS successfully corrected the defocus aberrations without showing a significant decrease in AO correction performance for increasing aberration strength. For aberrations of ±2 Diopter or stronger, the loop gain had to be adjusted in order to obtain convergence and the convergence time increased. If a similar dynamic range can be confirmed for other higher order aberrations in future work, it would be possible to perform P-WFS based AO imaging in a large part of the population. For persons with very large low order aberrations, AO correction after pre-correction with prescriptions lenses or glasses is a solution as has been demonstrated for volunteer V5* in Section S3.

## 5. IMAGING OF THE RETINA IN THE PERIPHERY WITH AN EXTENDED FIELD OF VIEW

In Figs. S8 and S9, representative AO-OCT image data recorded at the periphery of volunteer V4_R with scanning angles of 4° x 4° are shown underlining the capability of the P-WFS to support high resolution imaging at both small and large FoVs. The focus of the P-WFS based AO was set to the outer and inner retina (more specifically, the RPE and NFL layer) for the first and second volume, respectively. Both data sets are single AO-OCT volumes acquired with a 750 x 750 A-scan sampling.

The large FoV AO-OCT imaging data of the outer retina is presented in Fig. S6. In the en-face visulizations, the depth span from IS/OS to COST was divided into three integration ranges. The highly reflective spots appearing in the additional en-face image created over the outer segments (OS) of the cone photoreceptors (cf Fig. S6b) have been reported before [19] and probably correspond to defects in the packing arrangement of the cone outer segment discs. The clarity of Yellott's rings observed in the power spectra of the presented en-face images increases for the larger FoV as more cells are included in the analysis. At this eccentricity the IS/OS, COST and ROST layer show an elliptic Yellott's ring (cf. insets of Figs. S6a, S6c and S6d), which indicates that the packing density along a radial line from the fovea of the cone photoreceptors is higher than along the orthogonal direction. In contrast to this observation, the Yellott's ring of the RPE mosaic is perfectly circular (cf. inset of Fig. S6e), indicating a more homogenous packing arrangement of these cells. Structures of high sparsity or low signal strength, as for example the reflective spots on the OS layer or the vessel mesh of the choriocapillaris (CC), that might not be recognized as such with a small FoV are revealed in the large FoV images (cf. Figs. S6b and S6g, respectively).
In Fig. S7, the large FoV AO-OCT imaging data of the inner retina are displayed. In the averaged B-scan of Fig. S7f), the successful focus shift can be observed in the increased signal strength received from the inner retinal layers, especially the NFL. Lateral features of the inner retina can be observed at a high resolution over a larger imaging area in the en-face visualizations: The sparse distribution of the highly reflective circular and irregular structures providing evidence of Gunn's dots [20] and microglia [21] in the ILM (Fig. S7a) and the complex web of nerve fiber bundles of varying size dispersing across the NFL (Fig. S7b). The en-face slices extracted at the GCL (Fig. S7c), at the outer most part of the IPL (Fig. S7d) and at the bright band allocated with the OPL (Fig. S7e) show vessel patterns of the superficial, intermediate and deep capillary plexus, respectively. The areas between the vessels show patterns and irregularities, but the signal-to-noise ratio in the single-shot volumes is not sufficient to visualize cellular structures.

The large pupil size of the subject leads to a small depth of focus and the appearance of the intermediate and deep capillary plexus is slightly blurred. Due to the high numerical aperture provided by AO OCT, the vessel pattern appears with good contrast without the use of OCT angiography [22].

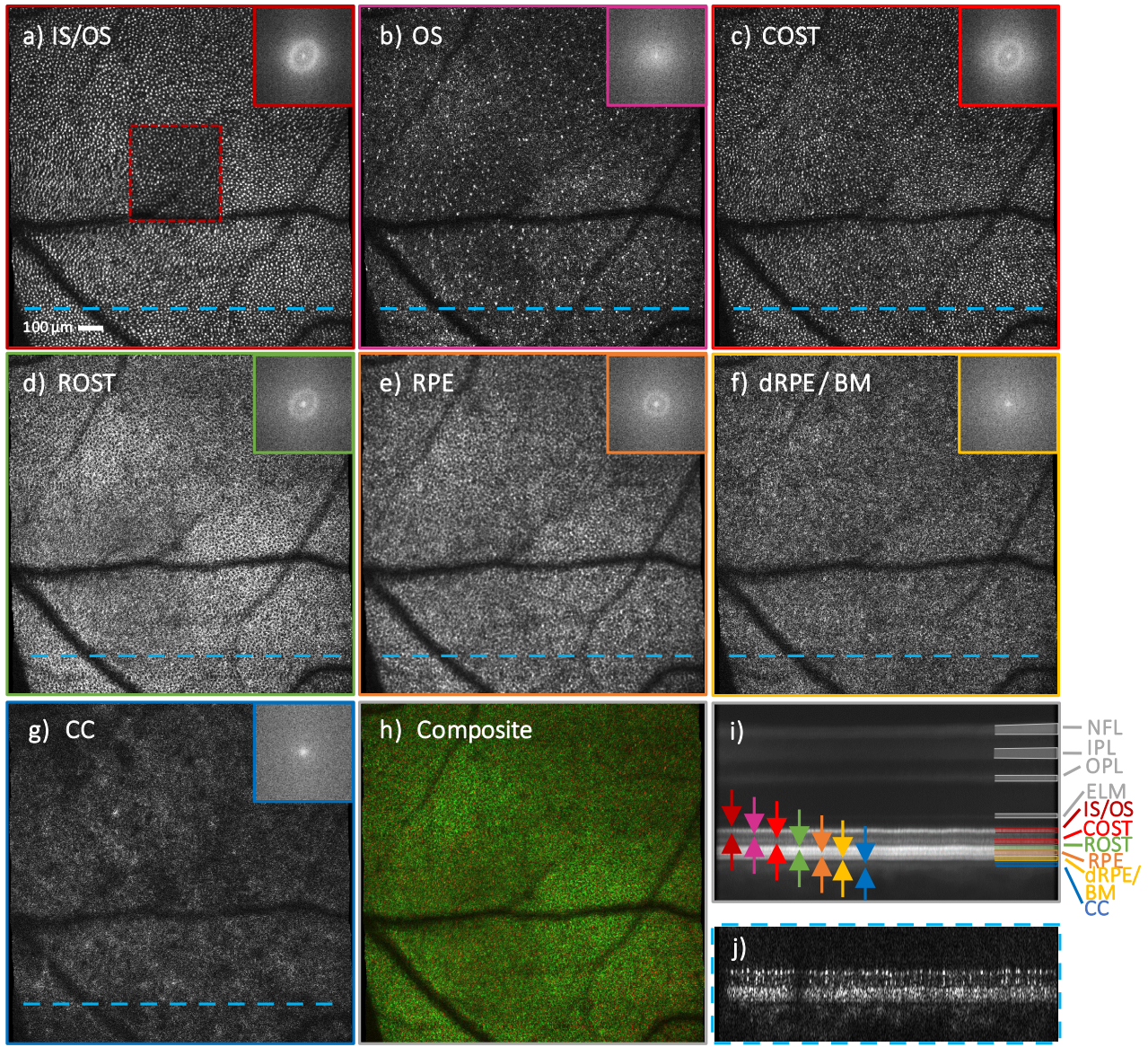

Fig. S8. Visualization of outer retinal layers using AO-OCT with the P-WFS at an extended FoV of 3.99° x 3.92°. The imaging location is 14° temporal / 6° superior in a healthy volunteer (29 years, male, right eye). The representative en-face images a)-h) are obtained from a single data volume by integration over layers in the outer retina and accompanied by their 2D FFTs in insets: a) junction between inner and outer segments of cone photoreceptors (IS/OS) with the small FoV image of Fig. 5. a) as inset, b) hyper-reflective spots in the cone outer segment layer (OS), c) cone outer segment tips (COST), d) rod outer segment tips (ROST), e) retinal pigment epithelium cells (RPE), f) presumptively the distal part of the RPE or Bruch's membrane (dRPE / BM), g) choriocapillaris (CC), h) composite false color image of COST (red) and ROST (green). The integration ranges are indicated by the color-coded arrows in the averaged B-scan i) which consists of a projection of the entire volume along the slow imaging axis. The location of the single B-scan in j) in the same data volume is marked by the dashed lines in the en-face images.

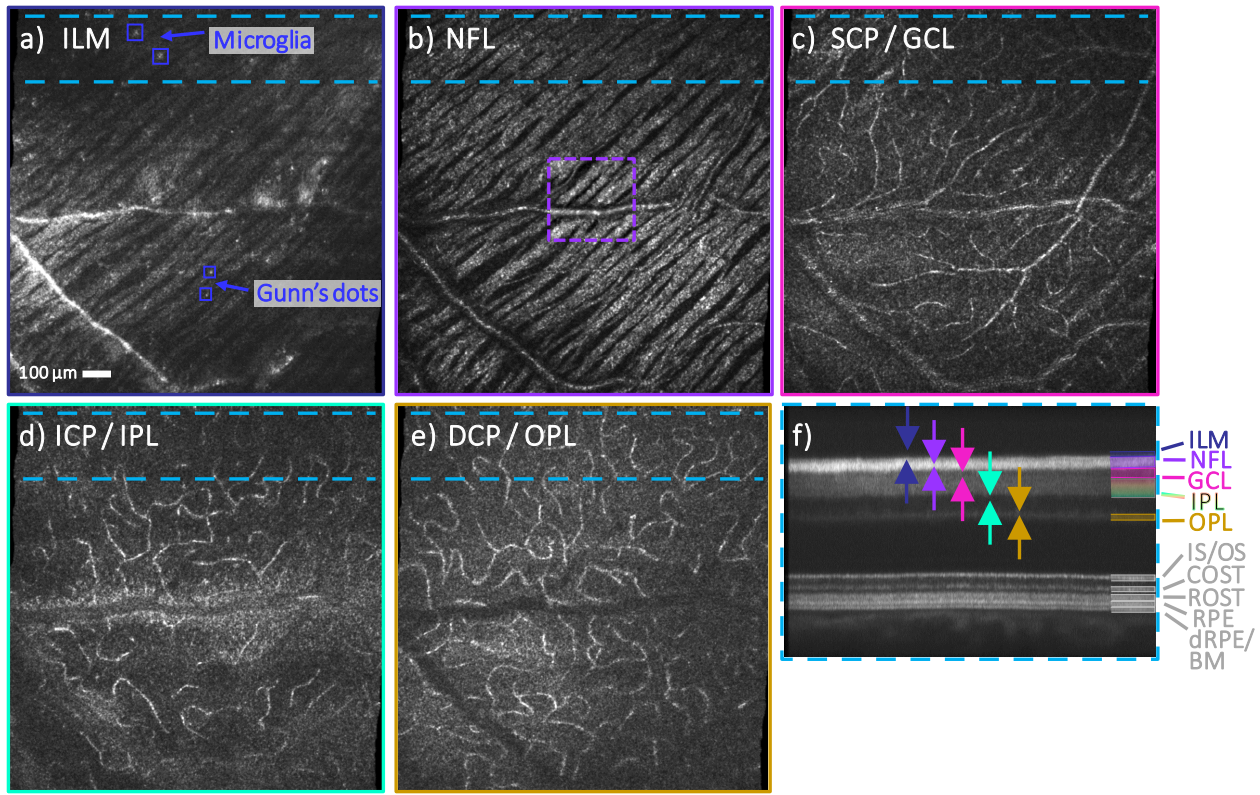

Fig. S9. Visualization of inner retinal layers using AO-OCT with the P-WFS at a large FoV of 3.81° x 3.92°. The imaging location 14° temporal / 6° superior in a healthy volunteer (29 years, male, right eye) and the focus was set to the nerve fiber layer. The representative en-face images a)-e) are extracted from a single data volume by depth integration in the inner retina over parts of or in vicinity of the following retinal layers: a) inner limiting membrane (ILM), b) nerve fiber layer (NFL) with the averaged, small FoV image of Fig. 7. b) as inset, c) superficial capillary plexus (SCP) in Ganglion cell layer (GCL), d) intermediate capillary plexus (ICP) in outer part of inner plexiform layer (IPL), e) deep capillary plexus (DCP) in outer plexiform layer (OPL). The integration ranges are indicated by the color-coded arrows in the averaged B-scan f) which consists of a projection of part of the volume along the slow imaging axis which is indicated by the dashed lines in the en-face images.